\documentclass[preprint,showpacs,preprintnumbers,amsmath,amssymb]{revtex4}
 \usepackage{graphicx}

\begin{document}
\title{Dependence of the TMCI threshold on the space charge tune shift}
\author{V. Balbekov}
\affiliation {Fermi National Accelerator Laboratory\\
P.O. Box 500, Batavia, Illinois 60510}
\email{balbekov@fnal.gov} 
\date{\today}

\begin{abstract}

Transverse mode coupling instability of a bunch with space charge 
is considered in frameworks of the boxcar model. 
Presented results demonstrate a monotonous growth of the TMCI threshold
at increasing space charge tune shift, 
and do not support the supposition that the monotony can be violated at 
a higher SC.

\end{abstract}

\maketitle
%

\section{INTRODUCTION}


Influence of the space charge (SC) on the transverse mode coupling instability (TMCI) 
has been investigated first by Blaskiewicz~\cite{BL1}.
The main conclusion which has be done by the author is that 
SC raises the TMCI threshold that is improves the beam stability.
However, several examples of a non-monotonic dependence of the TMCI parameters
on the SC tune shift have been represented in the paper as well.  
Similar results have been obtained also with help of numerical solution of  
equation of motion performed by the same author \cite{BL2}.  
The stabilizing effect has been certainly confirmed at relatively small SC 
tune shift. 

Special case of the dominant space charge is considered in papers \cite{BUR, BA1}. 
It is shown that the instability threshold grows up 
tending to infinity in this limiting case (``vanishing TMCI'').  
However, it was suggested in~\cite{BUR} that the threshold growth can 
cease and turn back 
if the tune shift $\,\Delta Q\,$ exceeds the synchrotron tune $\,Q_s\,$ 
by factor about ten or more.

The last statement has been supported in my recent paper~\cite{BA2}.
According to it, threshold of constant negative wake raises with SC 
up to $\,\Delta Q/Q_s\simeq 7\,$ but tends to 0 at larger space charge.
However, after an additional examination, I must revoke this statement
because of an insufficient accuracy of the numerical calculations was detected.
More exact solutions of the same dispersion equation is represented in this paper 
leading to  the conclusion that the TMCI threshold is a monotonously increasing  
function of the tune shift.
   

\section{Equation}


Following up Ref.~\cite{BA2}, I will consider transverse coherent oscillations of a
single bunch in the frameworks of boxcar model without chromaticity.
Dispersion equation for the coherent addition $\,\nu\,$ to the bare betatron 
tune $\,Q_0\,$ has been obtained there in form of the infinite continue 
fraction:
%
\begin{eqnarray}
 \nu-q + \frac{(q/3)^2 W_1}{1+\frac{(q/15)^2 W_1W_2}{1+\frac{(q/35)^2W_2W_3}
 {1+\dots\dots\dots}}} = 0,\qquad 
 W_n(\nu)=\sum_m\frac{|S_{n,m}|^2}{\nu-\nu_{n,m}}.
\end{eqnarray}
%
where $\,q\,$ is the normalized wake strength, $\,\nu_{n,m}\,$ 
and $\,S_{n,m}\,$ are the eigentunes and normalizing coefficients 
of the boxcar bunch without wake.
The last problem has been solved by Sacherer~\cite{SAH}, 
and a convenient description of the solution is available in \cite{BA2}.
Generally, $\,n=0,1,...$, and $\,m=n,n-2,...-(n-2),-n$.
In the conventional notation, $m$ is the multipole number whereas index 
$n$ is associated with number of radial mode.  
Figs.~1 and~2 are represented in this paper as the low index examples.
%
 \begin{figure}[t!]
 \includegraphics[width=100mm]{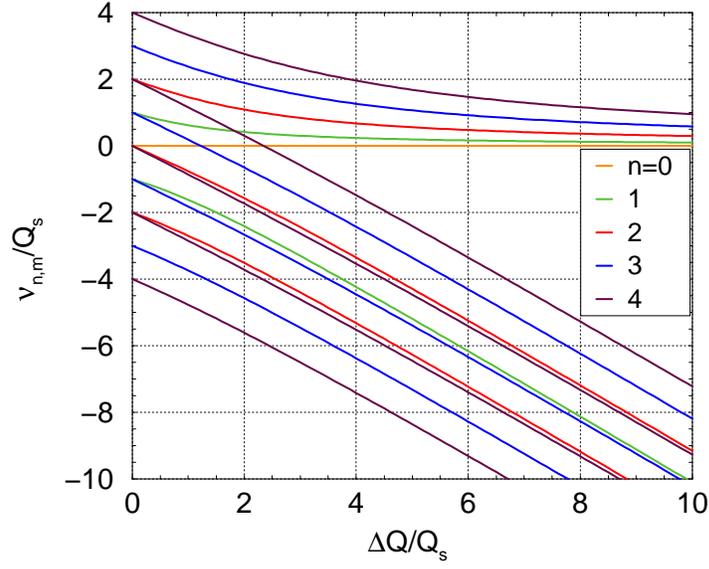}
 \caption{Eigentunes of the boxcar bunch without wake against space charge tune shift.}
 \vspace{5mm}
 \end{figure}
%
 \begin{figure}[t!]
 \includegraphics[width=100mm]{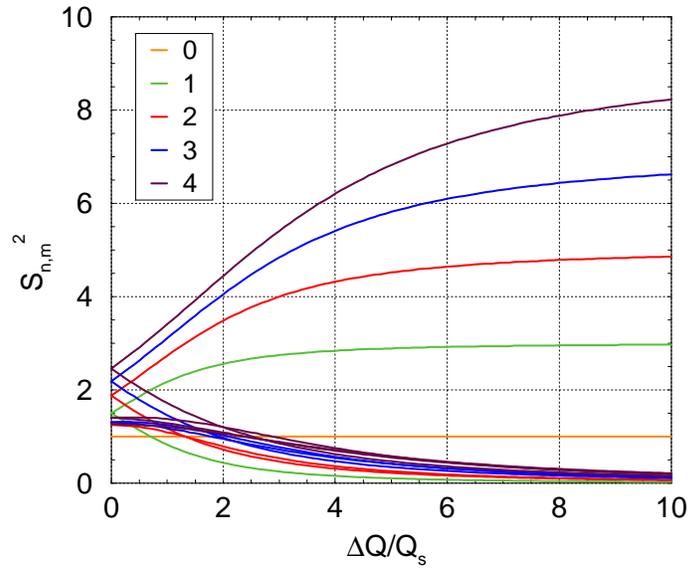}
 \caption{Normalized coefficients of the boxcar eigenfunctions against SC tune shift.} 
 \vspace{10mm}
 \end{figure}
%

If a restricted series of the modes is used, the approximate dispersion equation 
obtains the form  
%
\begin{eqnarray}
 T_{n_{\rm max}}(\nu)=0
\end{eqnarray}
%
with following recurrent relations:
%
\begin{eqnarray}
 T_n=T_{n-1}+T_{n-2}\frac{q^2W_{n-1}W_n}{(4n^2-1)^2}, \qquad n\ge2
\end{eqnarray}
%
and boundary conditions:
%
\begin{eqnarray}
 T_0=\nu-q,\qquad T_1 = \nu-q+\left(\frac{q}{3}\right)^2\frac{3(\nu+\Delta Q)}
 {\nu(\nu+\Delta Q)-Q_s^2}
\end{eqnarray}
%
First of them is actually the wake normalization condition to obtain the result 
$\,\nu=q\,$ in the lowest approximation.
%
 \begin{figure}[b!]
 \begin{center}
 \includegraphics[width=100mm]{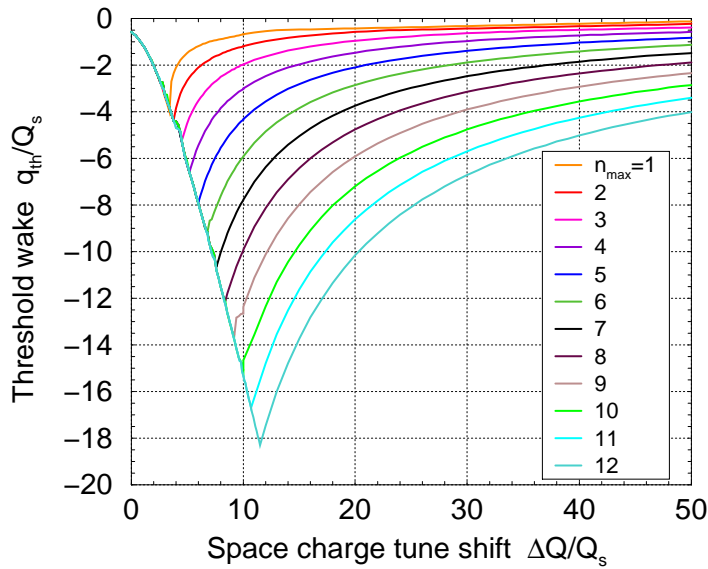}
 \end{center}
 \vspace{-10mm}
 \caption{Threshold value of constant negative wake against SC tune shift 
 in different approximations.}
 \end{figure}
%

Eq.~(2) is reducible to the algebraic equation of power 
$(n_{max}+1)(n_{max}+2)/2$.  
Generally, it can be solved numerically, and the TMCI threshold can be determined
by enumeration of real roots.
The results are represented in Fig.~3 at $\,n_{max}=1,\dots,12$.
A comparison with Figs.~7~and~8 of Ref.~\cite{BA2} demonstrates the full
coincidence of the curves at $\,n_{max}\le 6$. 
However, a drastic distinction appears at higher power leading us to revise 
the conclusions.

It has been declared in \cite{BA2} that the saturation occurs at $\,n_{max}= 6$,
and additional eigenvectors do not change the result even at very high value of 
$\,\Delta Q$.
Therefore decrease of the threshold at $\,\Delta Q/Q_s>7\,$ has been treated there 
as a real physical effect. 
However, this statement must be reconsidered now. 

It follows from Fig.~3 that the great convergence of different approximations 
occurs only in the falling part of the curves which corresponds to the increasing 
instability threshold.
It means also that the correct results can be ensured only at 
$\,\Delta Q/Q_s<n_{max}\,$.
The sequential decrease of the threshold should not be treated as a physical result
because of absence of the convergence.

The kink of the curves happens because of switching of the coalesced pairs 
which is illustrated by Fig.~4.
It is seen that the coalescence of the modes $\{n,m\}$
%
\begin{equation}
 \{0,0\}+\{1,-1\} 
\end{equation}
%
is responsible for the instability before the kink, and the combination
%
\begin{equation}
  \{n_{max},n_{max}\}+\{n_{max}-1,n_{max}-1\} 
\end{equation}
%
does it after the kink,
 with $\,n_{max}$ being dependent on the used approximation.
The mistaken assertion has be done in \cite{BA2} that (5,5)+(4,4) is the highest 
combination, and saturation appears after that.
This statement is not confirmed by the last calculations which are represented 
in Fig.~5.
The lower curves refer to the combination described by Eq.~(5) in different 
approximations. 
Very good convergence is seen in this part, so the addition on new basis 
eigenfunction merely prolongs the authentic area of instability.

A completely different situation appears in the upper part of the graph
where the highest tunes are represented being given by Eq.~(6) with 
different $\,n_{max}$. 
The lines merge at rather large $\,\Delta Q/Q_s\,$ which value depends on 
$\,n_{max}$.
However, an absence of the convergence compels to treat these regions as
the unphysical ones.


\section{Conclusion}


The more precise calculation demonstrate a monotonous growth of the TMCI threshold
of the negative wake at increasing space charge tune shift.
They do not support the supposition that the monotony can be violated at a higher 
SC tune shift. 
%
 \begin{figure}[t!]
 \hspace{-0mm}
 \begin{minipage}[t]{0.45\linewidth}
 \begin{center}
 \includegraphics[width=85mm]{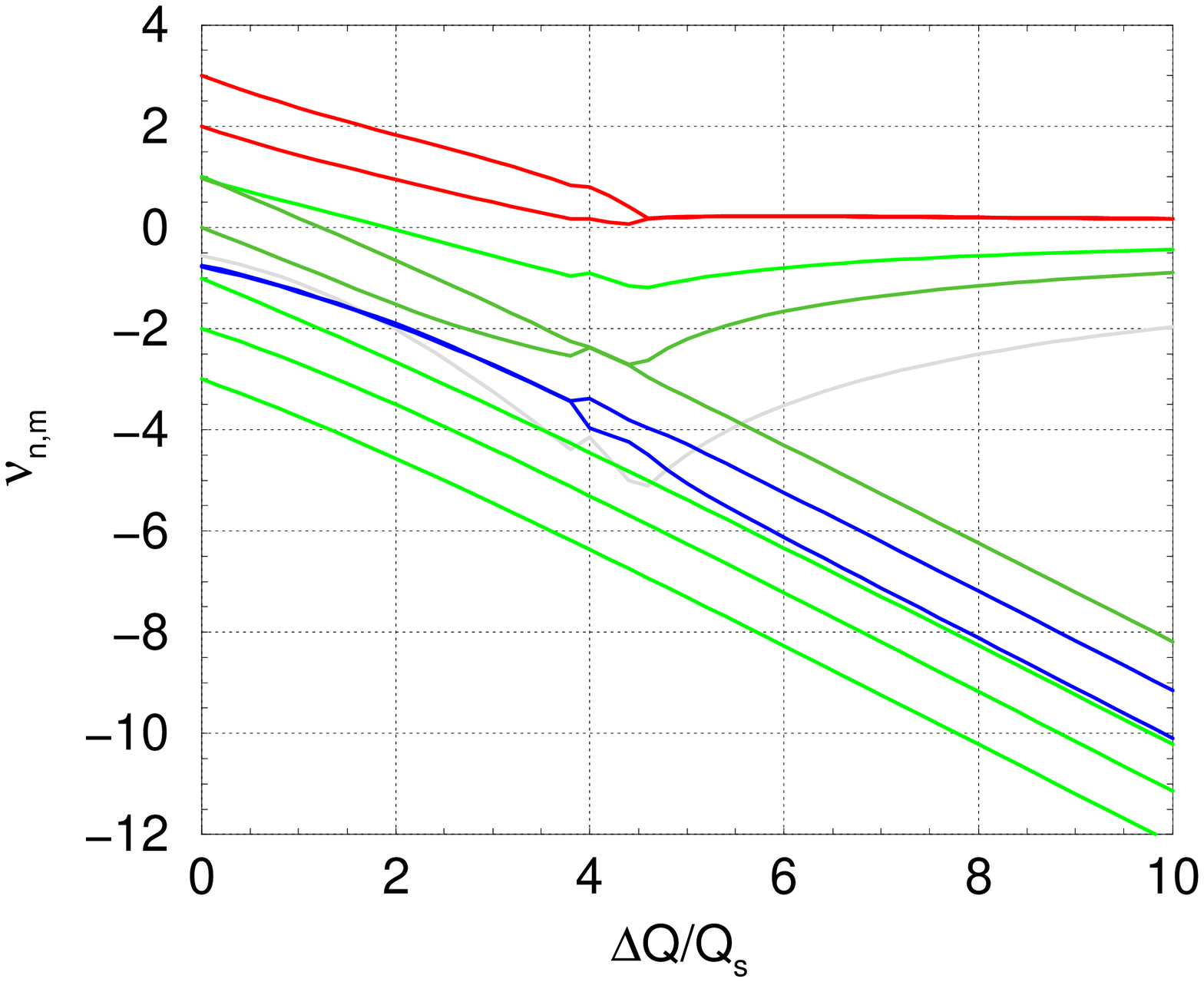}
 \end{center}
 \end{minipage}
 \hspace{5mm}
 \begin{minipage}[t]{0.45\linewidth}
 \begin{center}
 \includegraphics[width=85mm]{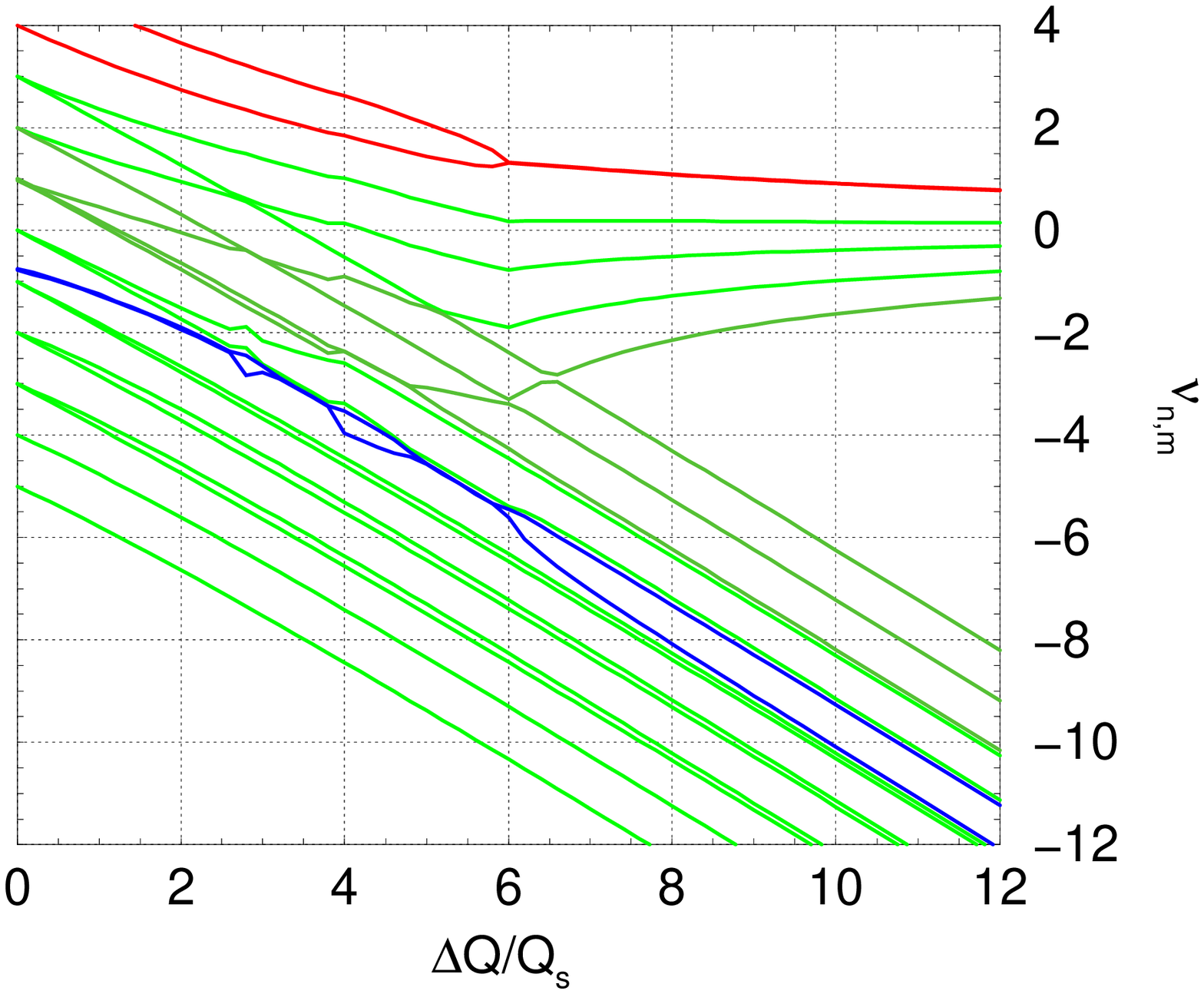}
 \end{center}
 \end{minipage}
 \caption{Boxcar bunch eigentunes against space charge tune shift at the 
 TMCI threshold.
 All eigentunes disclosed at given $\,n_{max}$ are shown, and the most important 
 of them (responsible for the instability) are marked by color. 
 Left: $\,n_{max}=3$, right: $\,n_{max}=5$.}
 \end{figure}
%
 \begin{figure}[b!]
 \begin{center}
 \includegraphics[width=100mm]{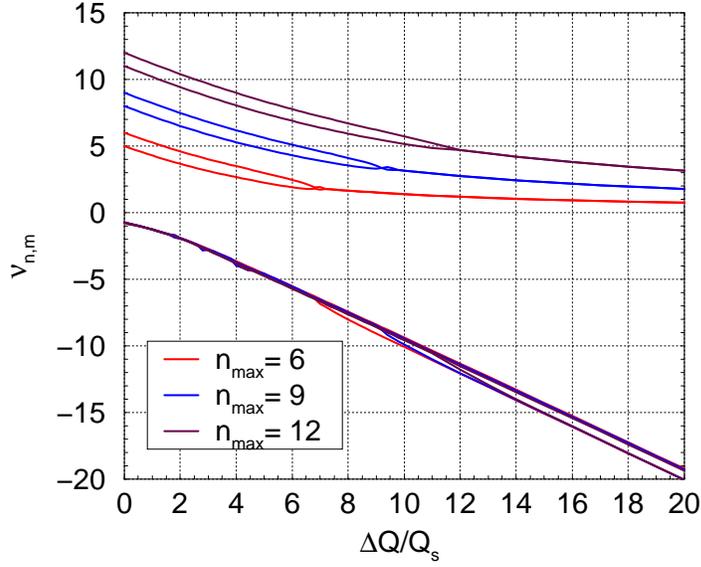}
 \end{center}
 \caption{Boxcar bunch eigentunes at the TMCI threshold against space charge 
 tune shift. 
 The most important tunes are shown.
 Upper lines: the higher tunes allowed by used set of eigenvectors.
 Lower lines: modes $\,\{0,0\}\,$ and $\,\{1,-1\}$.}
 \end{figure}
%


\end{document}